\documentclass{PoS}

\title{\bf Top-quark Decays to Higgs boson in  MSSM}

\ShortTitle{Top-quark Decays to Higgs boson in  MSSM}

\author{\speaker{K. Suxho}\\
        University of Ioannina\\
        E-mail: \email{csoutzio@cc.uoi.gr}}

\abstract{The decay of top-quark to light quarks and Higgs boson is an extremely rare process in the Standard Model (SM) framework. Any measurable signal for this process at the LHC is a smoking gun for new physics. In the framework of Minimal Supersymmetric Standard Model (MSSM) there is a possibility that the branching ratio for this process to reach values several orders of magnitude larger than the corresponding of SM. Having as a primary goal the search of Flavour Changing Neutral Currents (FCNC) processes that could be measurable at current or future LHC running, we present a calculation in the framework of $R$-parity conserving MSSM for the process $t\rightarrow q+h$, where $q$ is a light quark and $h$ is the Higgs boson\cite{DPRST}. Considering different constraints, and carefully looking at the cancellations that occur between different contributions, we deduce that there are two possible enhanced scenarios for this process:  
the first arises from the holomorphic SUSY breaking terms containing right handed up squark mass insertions and the second comes from the dominance of non-holomorphic SUSY breaking terms in combination with a light Higgs boson sector.}

\FullConference{18th International Conference From the Planck Scale to the Electroweak Scale \\
		25-29 May 2015\\
		Ioannina, Greece }

\begin{document}
\section{Introduction}
\setcounter{equation}{0}
The top-quark has been produced in large numbers at the LHC.  Therefore the LHC constitues an ideal place to search for top-quark decays. The two types of decays allowed from Lorentz invariance are: $t\rightarrow V+q$ and $t\rightarrow \phi+q$, where $q$ represents any light quark, $\phi$ represents the Higgs boson and $V$ represents the gauge bosons ($W$ and $Z$ gauge bosons, the photon $\gamma$ and the gluon g). The decay $t\rightarrow b+W$ is the dominant mode and well measured. The other decays, $t\rightarrow q+\gamma,\, t\rightarrow q+Z, \,t\rightarrow q+g,\, t\rightarrow q+h$ are extremely rare. We will concentrate on the last decay where in the final state light quarks (up and charm) are produced. In the (SM), the quark-scalar interaction originates from the Lagrangian\cite{PeskinShroeder}:
\begin{eqnarray}
\mathcal{L}_{SM}\supseteq -Y^{ij}_{u}\epsilon^{ab}\bar{Q}^{i}_{La}{H}^{\dagger} _{b}\,u^{j}_{R}-Y^{ij}_{d}\bar{Q}^{i}_{L}{H}\,d^{j}_{R}+h.c. ,
\end{eqnarray}
where $Y_{u},\,Y_{d}$  are general complex $3\times3$ matrices,
$u^{i}_{R}=\big(u_{R},c_{R},t_{R}\big)$ represents the right handed up-type quark fields which are $SU(2)$-singlets, $d^{i}_{R}=\big(d_{R},s_{R},b_{R}\big)$ represents the right handed down-type quarks fields which are $SU(2)$-singlets, $Q^{i}_{L}=\big(u_{L},d_{L}\big)^{T}, \big(c_{L},s_{L}\big)^{T}, \big(t_{L},b_{L}\big)^{T}$ correspond to left handed quark fields that are $SU(2)$-doublets and finally the Higgs field 
$H=\mathcal{U}\frac{1}{\sqrt{2}}\big(0,v+h\big)^{T}$ which is an $SU(2)$-doublet. In these relations, $T$ represents the matrix transposition and $\mathcal{U}$ represents a unitary transformation. By performing chiral transformations to quark fields the interaction terms take the form:
\begin{eqnarray}
\mathcal{L}_{SM}\supseteq-m^{i}_{u}\,\bar{u}^{i}_{L}\,u^{i}_{R}\,\Big(1+\frac{h}{v}\Big)+m^{i}_{d}\,\bar{d}^{i}_{L}\,d^{i}_{R}\,\Big(1+\frac{h}{v}\Big)+h.c.
\end{eqnarray}
It is important that the Higgs boson couples to quarks in a diagonal form. Therefore in the SM there are no $t\rightarrow q+h$ transitions at tree level!
The chiral transformations affect only the current:
\begin{eqnarray}
J^{\mu+}=\frac{1}{\sqrt{2}}\,\bar{u}^{i}_{L}\,\gamma^{\mu}\,\big(V_{CKM}\big)^{ij}\,d^{j}_{L}, 
\end{eqnarray}
which couples to $W^{+}$-field. Here the $V_{CKM}$ represents the Cabibbo-Kobayashi-Maskawa matrix. In this way the $t\rightarrow q+h$ transitions are induced \emph{only} by loop Feynman diagrams that contain these vertices. We should notice here that these diagrams contain the down-type quarks that have small mass differences and has been shown that $t\rightarrow q+h$ are sensitive to large mass differences of particles that are present in the loop. Taking these facts into account we can answer the question why the decay $t\rightarrow q+h$ is extremely rare in the SM. There are three reasons related to this suppresion:
 1) there is no tree level coupling, 2) the unitarity of $V_{CKM}$ gives an extra suppresion and finally, 3) the down-type quarks that enter in the loop have small mass differences. As a result of this suppresion, the branching ratio for top-quark decays to light quarks and Higgs boson is very small \cite{Eilamhewet}, \cite{MelePetrarka}, $\mathcal{B}(t\rightarrow u h)_{SM} \approx  4\times10^{-17},\, \mathcal{B}(t\rightarrow c h)_{SM} \approx  4\times10^{-14}$. This is far from the near future sensitivity of the LHC wich is about $\mathcal{O}\,(10^{-4})$. The LHC set some bounds on the $t\rightarrow q+h$ process. In order to estimate these bounds, we can write the Lagrangian in terms of two dimensionless quantities ($C^{(h)}_{L},\, C^{(h)}_{R}$), 
\begin{eqnarray}
\mathcal{L}&\supseteq & -C^{(h)}_{L}\,\bar{q}_{R}\,t_{L}\,h-C^{(h)}_{R}\,\bar{q}_{L}\,t_{R}\,h+h.c. ,
\end{eqnarray}  
which are called Wilson coefficients and contain all the information about the external momenta and the particle masses in interaction. The branching ratio as a function of Wilson coefficients has the following form:
\begin{eqnarray}
\mathcal{B}(t\rightarrow q h)=\frac{1}{1.39 GeV}\,\frac{m_{t}}{32 \pi}\Big(|C^{(h)}_{L}|^{2}+|C^{(h)}_{R}|^{2}\Big)\Big(1-\frac{m^{2}_{h}}{m^{2}_{t}}\Big)^{2} 
\approx  \frac{1}{4}\Big(|C^{(h)}_{L}|^{2}+|C^{(h)}_{R}|^{2}\Big).
\end{eqnarray}
Here, the constant $1.39\,GeV$ is the decay width of the process $t\rightarrow b+W$ and plays the role of a normalization factor, $m_{t}$ and $m_{h}$ correspond to top-quark and Higgs boson masses respectively. The current bound that LHC sets on these decays is: $\mathcal{B}(t\rightarrow q+h)\leq 0.79$ given by ATLAS \cite{Atlas}, and  $\mathcal{B}(t\rightarrow q+h)\leq 0.56$ given by CMS \cite{CMS}. This corresponds to an upper bound on $C_{L}$ and $C_{R}$: $|C_{L}|,\, |C_{R}|\lesssim 0.1$. The future bound that the LHC ($3000 fb^{-1}, 14\,\,TeV)$ will set on $t\rightarrow q+h$ is $\mathcal{B}(t\rightarrow q+h)\leq 2\times10^{-4}$ and this corresponds to $|C_{L}|,\, |C_{R}|\lesssim 0.01$. As we saw previously the SM prediction is at least ten orders of magnitude smaller than the LHC bound. Therefore, any signal for the $t\rightarrow q+h$ decay at the LHC means new physics beyond the SM.

\section{Cancellations and decoupling behaviour in the MSSM framework }
\begin{figure}[h!]
\begin{center}
\includegraphics[scale = 0.90]{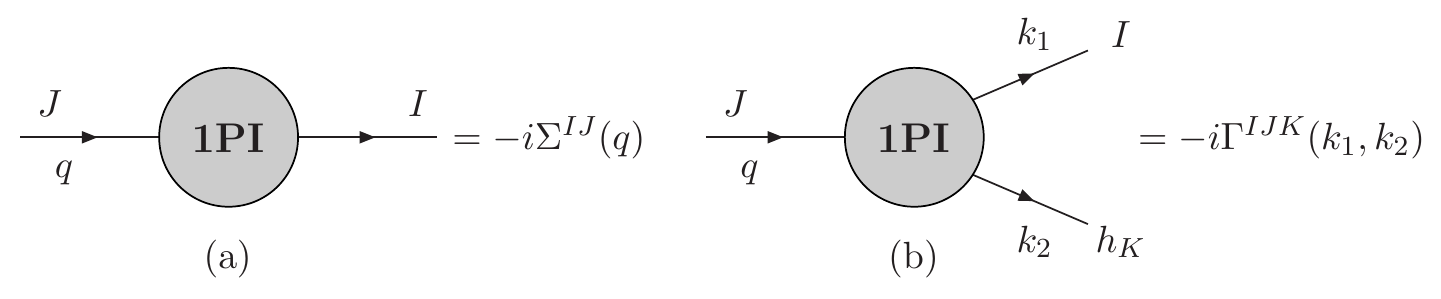}
\end{center}
\vspace{-0.5cm}
\caption{\sl The 1-particle irreducible diagrams that contribute to Wilson coefficients: (a) the self energy diagram, (b) the penguin diagram. Here $J$ and $I$ are flavour indices, $q$ is the momentum of the incoming top-quark and $k_{1}, \, k_{2}$ are the momenta of outgoing particles.}
\label{1PIdiag}
\end{figure}

\begin{figure}[h!]
\begin{center}
\includegraphics[scale = 0.60]{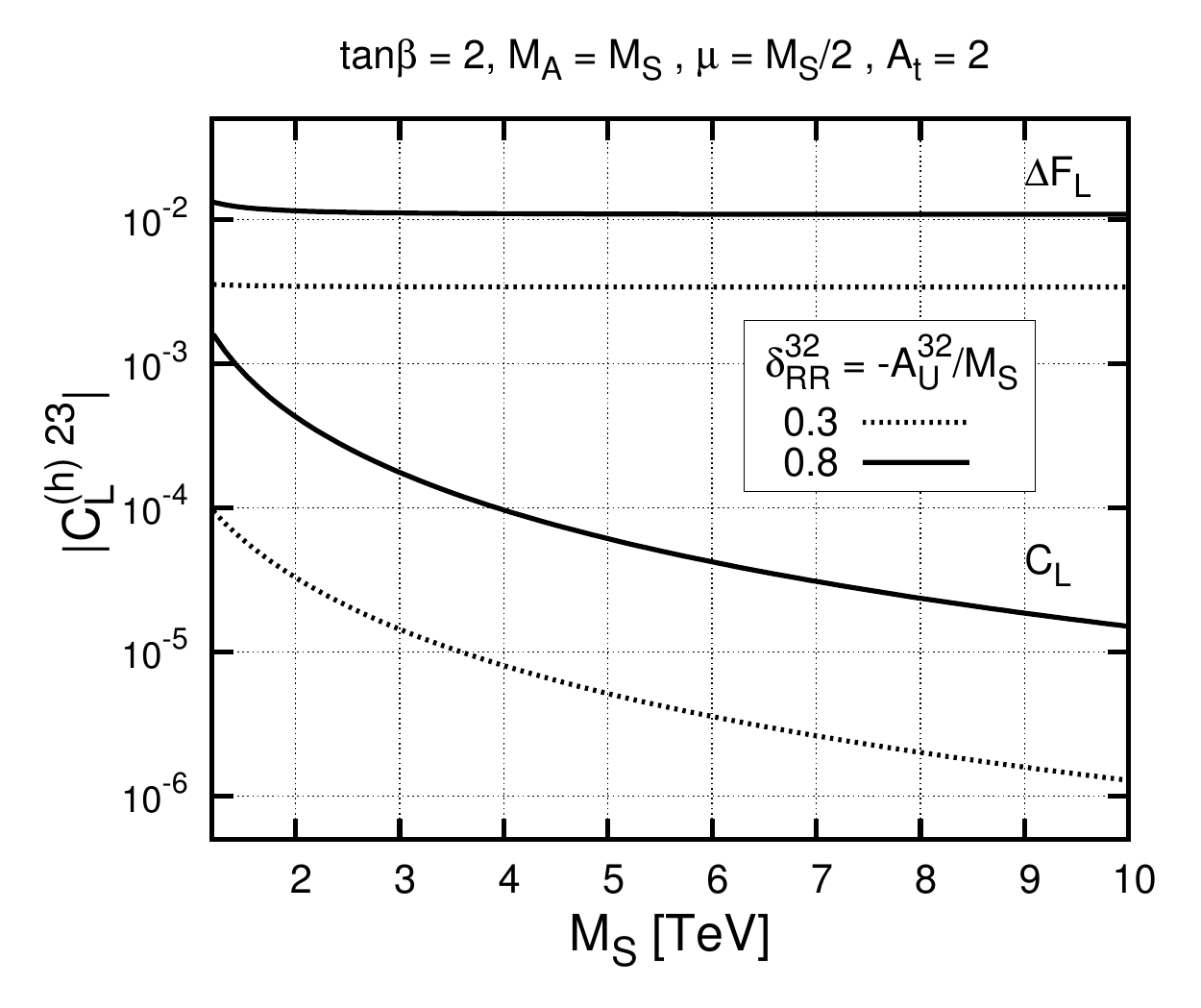}
\end{center}
\vspace{-0.5cm}
\caption{\sl The decoupling behaviour of Wilson coefficients. The contribution from the penguin diagram is shown at the top of the plot. It has a clear non-decoupling behaviour. When the contribution from the self-energy is added, the final result decouples as is shown at the bottom of the plot.}
\label{cancellationfig}
\end{figure}

\setcounter{equation}{0}
The SM prediction for the $t\rightarrow q+h$ decay is very pessimistic. Let us move to the MSSM framework and see what is different here. We are working in the R-parity conserving MSSM. In this case there are some optimistic conditions that allow a possible enhancement for the branching ratio  $\mathcal{B}(t\rightarrow q+h)$. Firstly, the Glashow-Iliopoulos-Maiani (GIM) mechanism that operates in the quark sector has a suppressing contribution  to the process in th SM framework. However this mechanism is not active in the squark interactions, allowing a possible enhancement. Secondly, instead of down-type quarks with small mass differences, now coloured scalars are present in the loops with potencially large mass differences. Depending on MSSM parameters, some authors claim that the branching ratio for the process $t\rightarrow q+h$ is of order $\mathcal{O}\,(10^{-4})$ \cite{GuashSola}, or $\mathcal{O}\, (10^{-5})$ \cite{CaoEilam} under some extra constrains (taking into account constraints from rare $B$-meson decays). We think that these predictions are very optimistic and we will show below that our best prediction is of order $\mathcal{O}\, (10^{-6})$. In order to investigate the source of a possible enhancement, we write the Lagrangian in the MSSM framework in terms of soft SUSY breaking matrices as following:
\begin{eqnarray}
{\cal L}_{\rm MSSM} & \supset & - \widetilde{Q}_{L}^{\dagger}
m_{Q_{L}}^{2} \widetilde{Q}_{L} - \widetilde{U}_{R}^{\dagger}
m_{U_{R}}^{2} \widetilde{U}_{R} - \widetilde{D}_{R}^{\dagger}
m_{D_{R}}^{2} \widetilde{D}_{R} \nonumber \\[1mm]
&+& \left ( H_{2}\: \widetilde{Q}_{L}\: A_{U}\: \widetilde{U}_{R} +
 H_{1} \: \widetilde{Q}_{L}\: A_{D}\: \widetilde{D}_{R}
 + \mathrm{H.c} \right ) \nonumber \\[1mm]
&+& \left ( H_{1}^{\dagger}\: \widetilde{Q}_{L}\:
 A^{\prime}_{U}\: \widetilde{U}_{R} +
 H_{2}^{\dagger} \: \widetilde{Q}_{L}\:
 A^{\prime}_{D}\: \widetilde{D}_{R} + \mathrm{H.c} \right ) \,, 
\end{eqnarray}
where $m_{Q_{L}}^{2}, \,m_{U_{R}}^{2},\, m_{D_{R}}^{2}$ are the soft SUSY breaking mass matrices, 
$A_{U}, \, A_{D}$ are the soft SUSY breaking trilinear matrices \cite{MisiakPokor} and
$A^{\prime}_{U},\, A^{\prime}_{D}$ are the non-holomorphic soft SUSY breaking trilinear matrices \cite{HallRandall},\cite{Borzumati}. In general, the structure of soft breaking terms is not trivial. For this reason, the quark and squark sector is not possible to be diagonalised at the same time in the same flavour basis. On the other hand, taking into account kaon, charm and $B$-physics experiments, a fully generic structure of these matrices has been excluded. The only exception of this exclusion is the right handed up-squarks mass matrix as well as the trilinear SUSY breaking matrices $A_{U}$ and $A^{\prime}_{U}$. In order to calculate the Wilson coefficients, we use the 1-particle irreducible diagrams which are shown in Fig. \ref{1PIdiag}.
\vspace{1.0cm}
\vspace{0.5cm}
Under the assumption that the final quark has a very small mass ($m_{I} = m_{u}(m_{c})\to 0$), we find for the Wilson coefficients the following expression:
\begin{equation}
C_{L}^{(h)\, IJ} \ = \ \Delta F_{L}^{(h)\, IJ} \ - \ \frac{1}{v}\, 
\left (\frac{\cos\alpha}{\sin\beta} \right ) \, \Sigma_{mL}^{IJ}(0)\;,
\label{WilCoeff}
\end{equation}
where in our case $I=1,\,2$ and $J=3$ and $\alpha$ is an angle related to rotations in the Higgs fields sector. In eq. \ref{WilCoeff}, the first term comes from the penguin diagram and the second term comes from the self energy diagram. From eq. \ref{WilCoeff} we can see that there is no $\tan{\beta}$ enhancement. Also there are huge cancellations between the two contributions. In fact the Wilson coefficients are functions of mass insertions $\delta^{IJ}$ which are defined in the following way \cite{MisiakPokor}, \cite{Gabbiani}:
\begin{eqnarray}
\delta_{{X}}^{JI} = \frac{(m_{X}^{2})^{JI}}{\sqrt{(m_{X}^{2})^{II}\, 
(m_{X}^{2})^{JJ}}} \,\,,
\end{eqnarray}
and contain the contribution from the non-diagonal terms of mass matrices. In general the Wilson coefficients as functions of mass insertions have the form:
\begin{eqnarray}
C_{L,R}^{(h)} \ \approx \ \frac{\alpha_{s}}{4\pi}\, \left
(\frac{m_{\tilde{g}}}{M_{S}} \right)\,f(\delta_{LR}^{32}...).
\label{coeffdelta}
\end{eqnarray}
In fact all particle corrections have been taken into account, but due to the strong coupling $\alpha_{s}$, the diagram that contains the gluino in the loop is the dominant. 
In eq. \ref{coeffdelta}, $m_{\tilde{g}}$ is the gluino mass and $M_{S}$ is a SUSY mass scale. The plot in Fig. \ref{cancellationfig}       is a graphical confirmation of cancellations that take place between contributions from the penguin and self energy diagrams and shows that the final result has a decoupling behaviour.
It is clear that the contribution from the penguin diagram (the two horizontal lines at the top) clearly has a non-decoupling behaviour, but when we add the self energy contribution the non-decoupling terms cancel each other and the final result has a decoupling behaviour (the two lines at the bottom). 
 After cancellations between the self energy and penguin contributions, the remaining corrections are proportional to $m^{2}_{t}/M^{2}_{S}$ in the best case and are collected in the following list \footnote{In order to arrive at this result we used an expansion approximation of mass matrices using the notion of divided differences \cite{divdif}. An analytic, automatic method to calculate this expansion using Mathematica, has been developed recently \cite{RosiekMathematica}.}:
\begin{eqnarray}
&\sim &\: A'^{JI}_U\: \frac{\cos(\alpha-\beta)}{\sin{\beta}} 
\:\:\times \:{\cal{O}}\left(\frac{1}{M_S}\right)\qquad \,\,\,\,\,
\sim \: \delta_{RR}^{JI}\left(
\frac{\cos{\alpha}}{\sin{\beta}}\right) \times
{\cal{O}}\left(\frac{m_t^2}{M_S^2}\right)\nonumber \\ 
&\sim &\:\mu^\star \delta_{RR}^{JI}\: \frac{\cos(\alpha-\beta)}{\sin{\beta}}
\:\times \:{\cal{O}}\left(\frac{1}{M_S}\right) \qquad
 \sim \:\sum_{A=1}^3\delta_{RL}^{JA}\delta_{LR}^{AI}
\:\:\left(\frac{\cos{\alpha}}{\sin{\beta}}\right)
\times {\cal{O}}(1)\nonumber \\
&\sim &\: \delta_{LR}^{JI}\:\left(\frac{\cos{\alpha}}{\sin{\beta}}\right)\:
\times {\cal{O}}\left(\frac{m_t}{M_S}\right) \qquad \,\,\,\,\,\,\,\,\,\,\,\,\,\,
\sim\: \delta_{LR}^{JJ}\delta_{RR}^{JI}\: 
\left(\frac{\cos{\alpha}}{\sin{\beta}}\right)\:
\times {\cal{O}}\left(\frac{m_t}{M_S}\right) \nonumber \\
&\sim &\: \sum_{A,B=1}^{3}\: \delta_{LR}^{JA}\,\delta_{RL}^{AB}
\,\delta_{LR}^{BI}
\left(\frac{\cos{\alpha}}{\sin{\beta}}\right)
\times {\cal{O}}\left(\frac{M_S}{m_t}\right)\;.
\label{remnants}
\end{eqnarray}
Here $\delta$-s are the mass insertions and $\delta_{RR}\sim \mathcal{O}(1)$ and $\delta_{LR},\,\delta_{RL}\sim m_{t}/M_{S}$. In fact the $\delta$-s are $3\times 3$ matrices and by definition the diagonal terms of $\delta_{LL}$ and $\delta_{RR}$ are zero. The list in eq. \ref{remnants} will guide us to search for possible enhanced scenarios.
\begin{figure}[h!]
\begin{center}
\includegraphics[scale = 0.75]{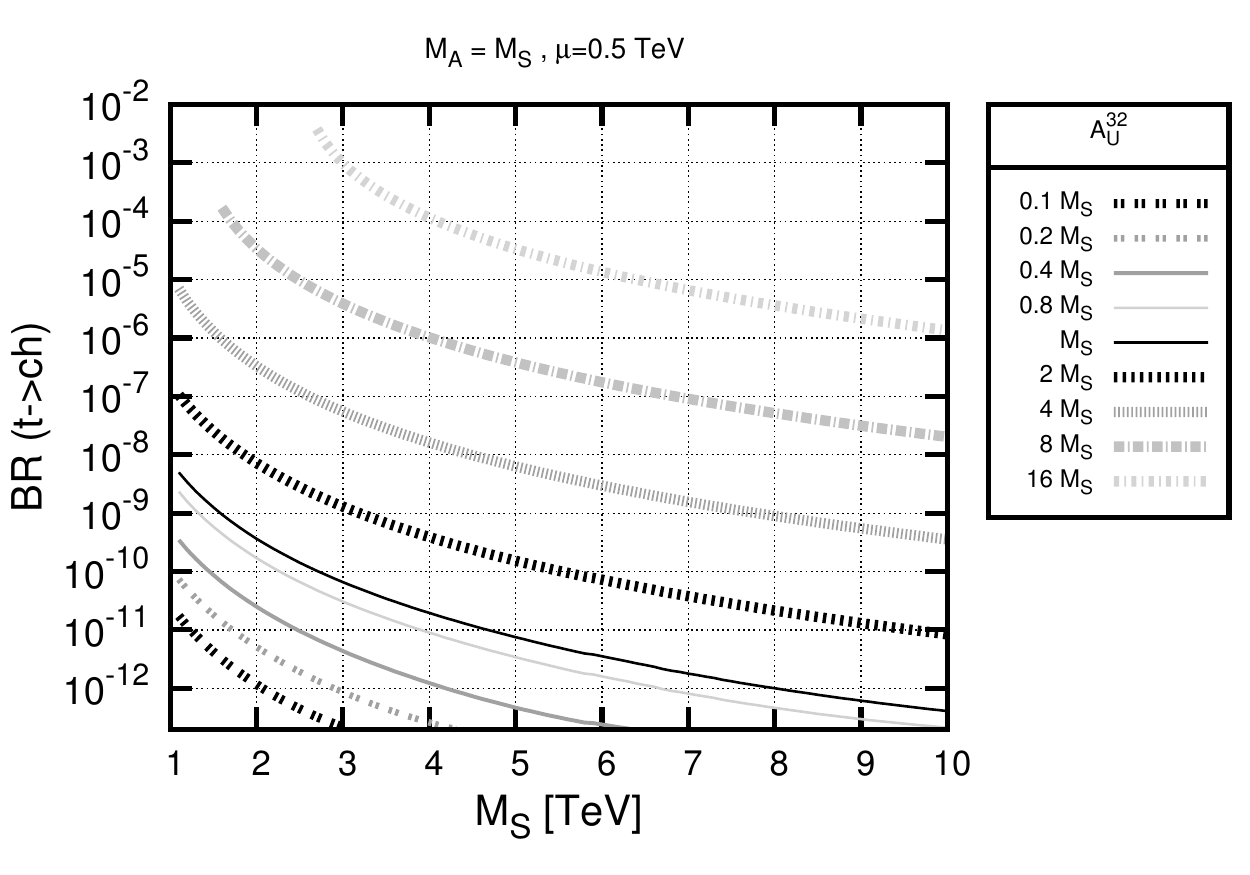}
\end{center}
\vspace{-0.5cm}
\caption{\sl The dependence of $\mathcal{B}(t\rightarrow c+h)$ from the SUSY mass scale $M_{S}$ for various values of $A^{32}_{U}/M_{S}$. We assume a uniform scaling $m_{\tilde g} = M_{A} = M_{S}$  and $2\leq \tan{\beta} \leq4$. The decoupling behaviour is evident for each value of $A^{32}_{U}/M_{S}$.
}
\label{decoupling1}
\end{figure}
A parameter which will provide a possible enhancement is $\delta^{32}_{LR}$. In Fig. \ref{decoupling1}, the dependence of $\mathcal{B}(t\rightarrow c+h)$ from the SUSY mass scale $M_{S}$ is shown, for various input values of the ratio $A^{32}_{U}/M_{S}\sim\delta^{32}_{LR}$. We can see that for each value of this ratio there is a decoupling behaviour. More interesting are the plots that correspond to values of $\delta^{32}_{LR}> 1$. But how realistic are these plots?
 It is clear from the plot that for $\delta^{32}_{LR}> 8$ the branching ratio can become observable at the LHC ($\mathcal{B}(t\rightarrow c+h)\geq 10^{-4}$). Such large values for $A^{32}_{U}$ are not allowed because they can trigger unwanted Charge and Colour Breaking (CCB) minima \cite{CasasDimop}. Under some conditions for the scalar fields we find the following constrain for the $A^{32}_{U}$ parameter\cite{DPRST}:
\begin{equation}
|A_{U}^{32}|^{2} \ \lesssim \ Y^{2}_{t} \,(m_{H_{2}}^{2} +
m_{\tilde{t}_{L}}^{2} + m_{\tilde{c}_{R}}^{2} + \mu^{2}), 
\end{equation} 
 and assuming a common squark and Higgs mass scale, this results in $A^{32}_{U}\leq \sqrt{3}\,M_{S}$. After taking into account this constrain we deduce that $\mathcal{B}(t\rightarrow c+h)\leq 10^{-7}$ and this is far from the near future LHC sensitivity. 
 
\begin{figure}[h!]
\begin{center}
\includegraphics[scale = 0.70]{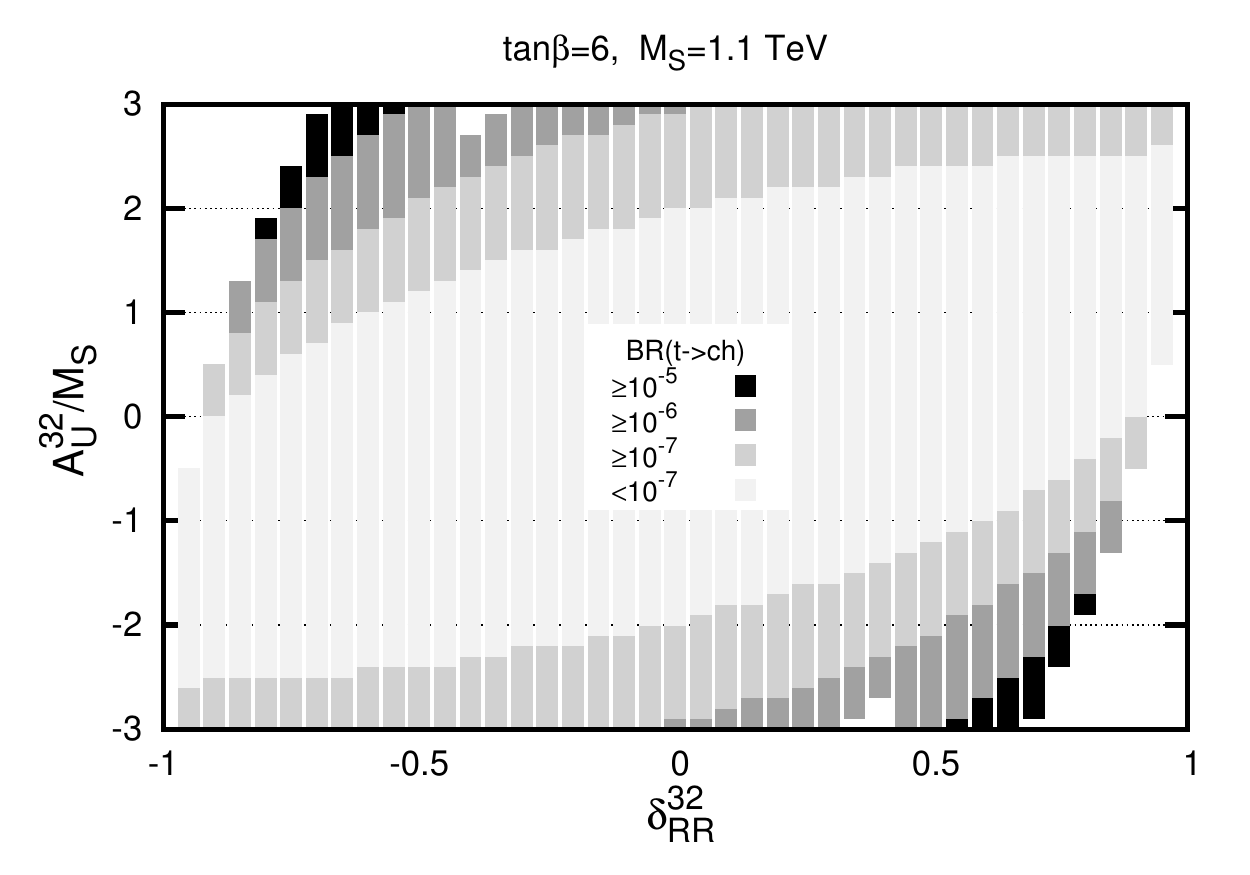}
\end{center}
\vspace{-0.5cm}
\caption{\sl Parametric plot of $\mathcal{B}(t\rightarrow c+h)$ on the $\delta^{32}_{RR},\,\,A^{32}_{U}/M_{S}$ plane. In the allowed region, the branching ratio can not excced values of $10^{-5}$, wich is smaller than the LHC sensitivity.}
\label{deltaRRanddeltaLR}
\end{figure}

 In a more general case both parameters $A^{32}_{U}/M_{S}$ and $\delta^{32}_{RR}$ can be present simultaneously. In Fig. \ref{deltaRRanddeltaLR}  we plot $\mathcal{B}(t\rightarrow c+h)$ on $\delta^{32}_{RR},\,\,A^{32}_{U}/M_{S}$ plane varying $A^{32}_{U}$ in a region to avoid potencial CCB bounds. We can see that the two parameters can interfere constructively or destructively along the two diagonal directions. However, in the most optimistic case the branching ratio can not exceed values of order $10^{-5}$.
\newpage
\vspace{-0.2cm}

\begin{figure}[h!]
\begin{center}
\includegraphics[scale = 0.68]{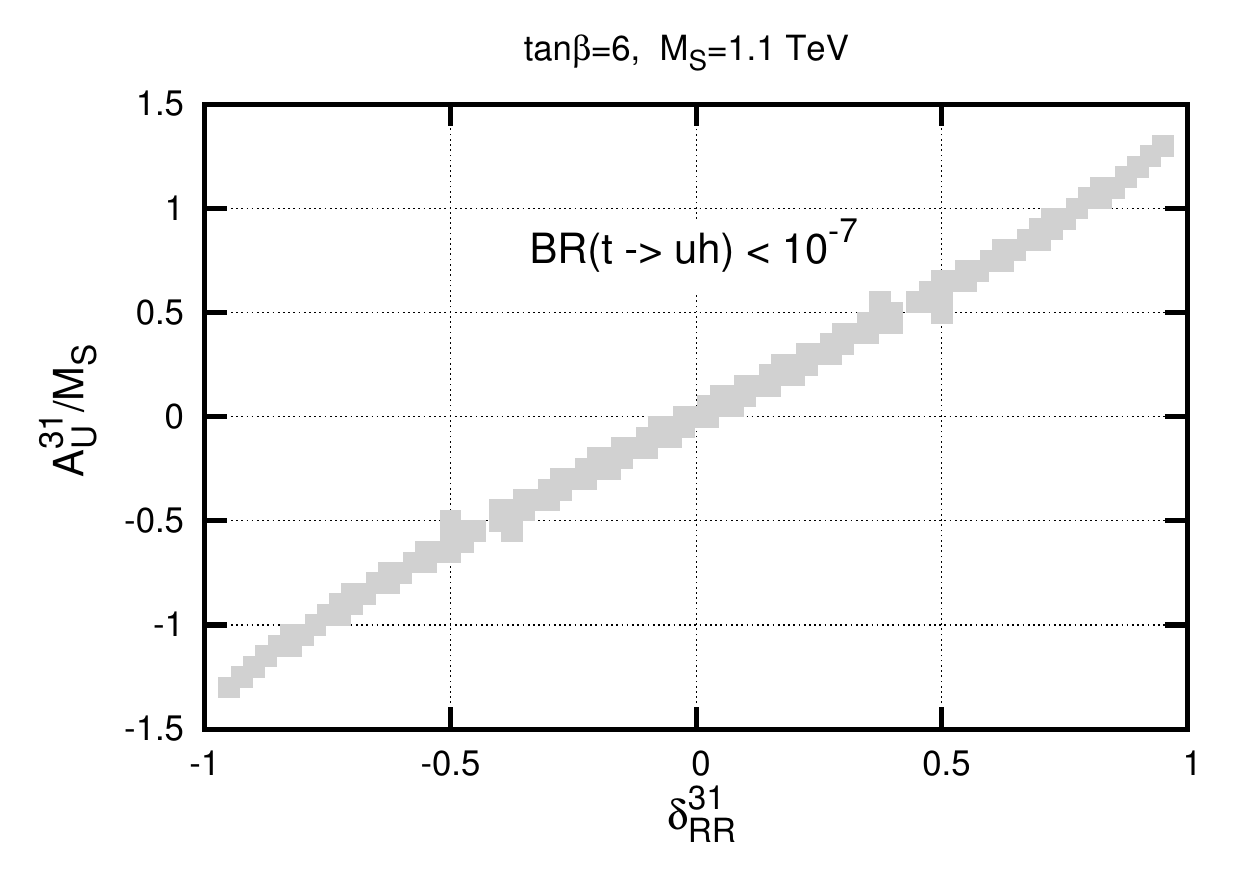}
\end{center}
\vspace{-0.7cm}
\caption{\sl Parametric plot of $\mathcal{B}(t\rightarrow u+h)$ on the $\delta^{31}_{RR},\,\,A^{31}_{U}/M_{S}$ plane. The branching ratio has an extra suppression from the neutron electric dipole moment \cite{divdif}.}
\label{neutronedm}
\end{figure}

In Fig. \ref{neutronedm} we present results for $\mathcal{B}(t\rightarrow u+h)$ on $\delta^{31}_{RR},\,A^{31}_{U}/M_{S}$ plane. 
The important difference from the previous figure is the fact that the $A^{31}_{U}$ and $\delta^{31}_{RR}$ paramters are highly constrained by experimental bounds from the neutron electric dipole moment. As previously we find in the best case $\mathcal{B}(t\rightarrow u+h)\leq 10^{-7}$ which is unobservable at the LHC.
\vspace{-0.2cm}

\begin{figure}[h!]
\begin{center}
\includegraphics[scale = 0.61]{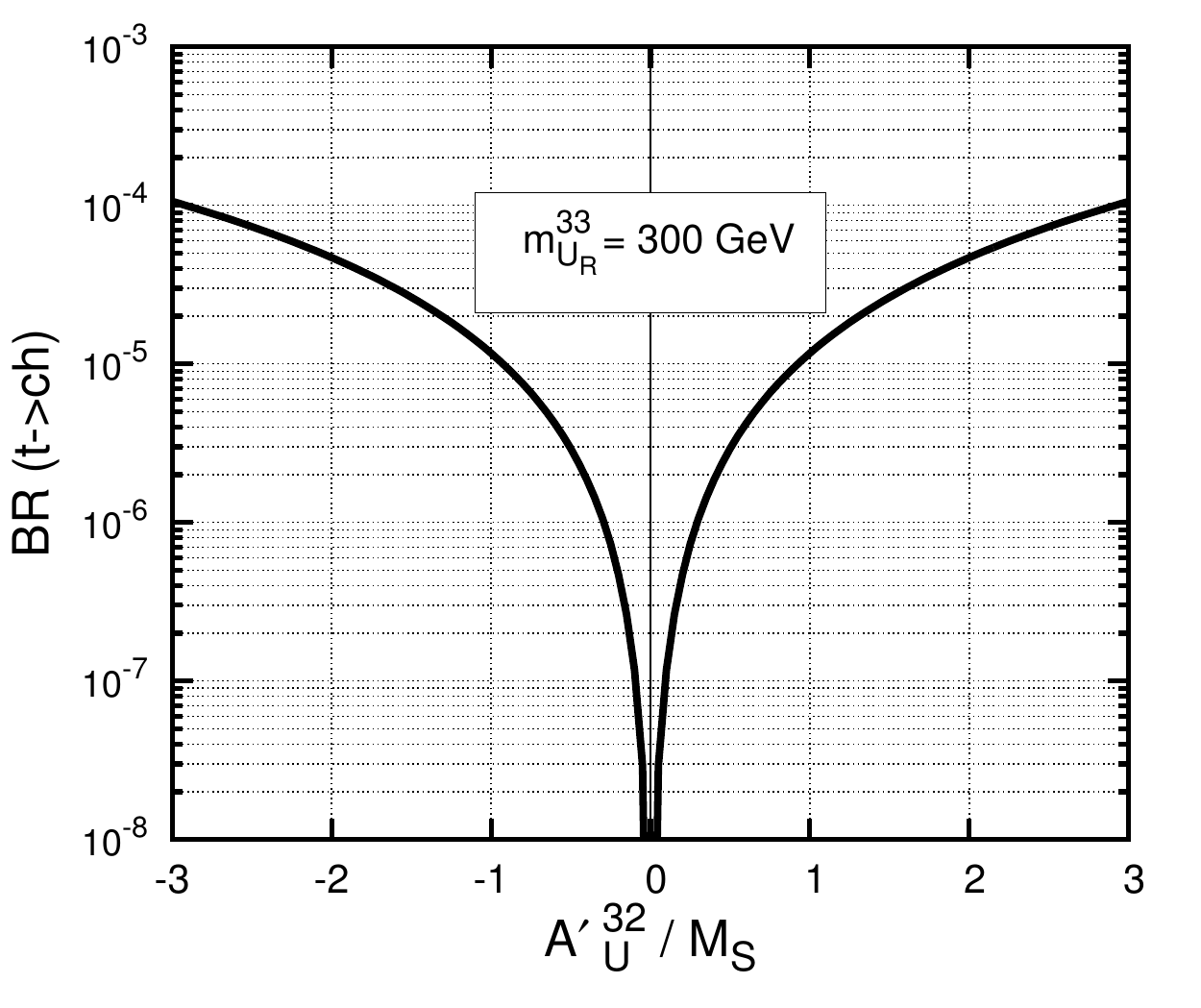}
\end{center}
\vspace{-0.7cm}
\caption{\sl The dependence of $\mathcal{B}(t\rightarrow c+h)$ from the non-holomorphic parameter $A'^{32}_{U}/M_{S}$ considering a light Higgs sector. 
Althought it seems that the branching ratio can reach values measurable at the LHC, 
other data disfavour this scenario.
}
\label{nonholomorphic}
\end{figure}

Finally, there is also an alternative enhancement scenario which requires a light Higgs sector and also a considerable contribution from the non-holomorphic trilinear coupling $A^{'}_{U}$ \cite{Drees}. We assume a light Higgs sector ($M_{A}\sim M_{Z}$) in order terms proportional to $\cos(\alpha-\beta)$ in eq. \ref{remnants} to be enhanced. According to \cite{Drees} we follow the scenario where the heavy Higgs boson is the one observed at the LHC and the light is about $95-100\,GeV$. In Fig. \ref{nonholomorphic} we present a plot of $\mathcal{B}(t\rightarrow c+h)$ as a function of non-holomorphic parameter $ A'^{32}_{U}/M_{S}$. We observe that for large values of $ A'^{32}_{U}/M_{S}$ parameter, the branching ratio can reach observable values at the LHC. However this scenario is disfavoured by LHC data \cite{CMS} and we present the plot only for complementarity reasons.
\newpage
\section{Conclusions}
As discussed in the Introduction the branching ratio for the processes $t\rightarrow q+h$ is extremely small in the SM framework, far from the current and near future LHC sensitivity. However, for this processes, the MSSM framework provides new possibilities. Taking into account the cancellations that occur between different contributions and a variety of experimental constraints and other related to CCB constraints, we find $\mathcal{B}(t\rightarrow q+h)\leq 10^{-6}$. There are huge cancellations between the self energy and penguin contributions and the remaining terms are in the best case proportional to $m^{2}_{t}/M^{2}_{S}$. We consider the effects of NLO-QCD corrections due to the SUSY loop induced chromomagnetic dipole operator and the running of
operators from the SUSY scale $M_{S}$ to the top-quark scale. Also we provide an analytical and detailed calculation of cancellations and decoupling using a common scheme for both universal and
hierarchical squark mass structures. As an alternative scenario about the enhancement of the final effect, we investigate how the branching ratio is affected by the non-holomorphic SUSY breaking terms $A^{'}_{U}$. Finally, we have encoded all our calculations into a publicly available
code (SUSY{\_}FLAVOR code \cite{codeRosiek}), where a variety of up-to-date experimental constraints has been also
included. All the numerical calculations and plots have been performed with the use of {SUSY{\_}FLAVOR} code \cite{codeRosiek}, \cite{codeRosiek10}, \cite{codeRosiek12}.

\section{Acknowlegments}
The author wishes to thank A.Dedes, M.Paraskevas, J.Rosiek, and K.Tamvakis, for valuable discussions. 
This research has been
co-financed by the European Union (European Social Fund - ESF) and Greek national funds
through the Operational Program "Education and Lifelong Learning" of the National Strategic Reference Framework (NSRF) - Research Funding Program:  ARISTEIA -
Investing in the society of knowledge through the European Social Fund.

\end{document}